\documentclass[aps,amsmath,prd]{revtex4}

\usepackage{graphicx}

\def\l{\left}
\def\r{\right}
\def\be{\begin{equation}}
\def\ee{\end{equation}}
\def\bea{\begin{eqnarray}}
\def\eea{\end{eqnarray}}

\begin{document} \title{
Searching for Gravitational Waves from the Inspiral of Precessing Binary
Systems:

Problems with Current Waveforms.}

\author{Philippe Grandcl\'ement}
\email[]{PGrandclement@northwestern.edu}
\affiliation{Northwestern University, Dept.\ of Physics \& Astronomy, 2145 
Sheridan Road, Evanston 60208, USA}

\author{Vassiliki Kalogera}
\email[]{vicky@northwestern.edu}
\affiliation{Northwestern University, Dept.\ of Physics \& Astronomy, 2145 
Sheridan Road, Evanston 60208, USA}

\author{Alberto Vecchio}
\email[]{av@star.sr.bham.ac.uk}
\affiliation{School of Physics and Astronomy, 
The University of Birmingham, Edgbaston, Birmingham B15 2TT, UK}

\date{November 22, 2002}

\begin{abstract}

We consider the problem of searching for gravitational waves emitted
during the inspiral phase of binary systems when the orbital plane
precesses due to relativistic spin-orbit coupling. Such effect takes place
when the spins of the binary members are misaligned with respect to the
orbital angular momentum. As a first step we assess the importance of
precession specifically for the first-generation of LIGO detectors. We
investigate the extent of the signal-to-noise ratio reduction and, hence,
detection rate that occurs when precession effects are not accounted for
in the template waveforms. We restrict our analysis to binary systems that
undergo the so-called simple precession and have a total mass $\lesssim
10\,M_{\odot}$.  We find that for binary systems with rather high mass
ratios (e.g., a 1.4\,M$_\odot$ neutron star and a 10 M$_{\odot}$ black
hole) the detection rate can decrease by almost an order of magnitude.
Current astrophysical estimates of the rate of binary inspiral events
suggest that LIGO could detect at most a few events per year, and
therefore the reduction of the detection rate even by a factor of a few is
critical. In the second part of our analysis, we examine whether the
effect of precession could be included in the templates by capturing the
main features of the phase modulation through a small number of extra
parameters. Specifically we examine and tested for the first time 
the 3-parameter family
suggested in \cite{Apost96}. We find that, even though these ``mimic''
templates improve the detection rate, they are still inadequate in
recovering the signal-to-noise ratio at the desired level. We conclude
that a more complex template family is needed in the near future, still
maintaining the number of additional parameters as small as possible in
order to reduce the computational costs.
 \end{abstract}

\pacs{}
\maketitle

 \section{Introduction}

With the international network of ground-based gravitational wave (GW)
detectors LIGO \cite{Abram92}, VIRGO \cite{Caron97}, GEO600
\cite{Danzm95}, TAMA \cite{Tagos01} coming online, the need for accurate
source modeling, which can guide the construction of optimal data analysis
strategies and therefore maximizes the detection efficiency is becoming
increasingly pressing. The expected signals will be so weak compared to
the intrinsic noise of the detectors, that one relies on a number of
different data-processing techniques for signal extraction and detection,
followed by search for coincidences in two or more instruments.
Matched-filtering provides the optimal linear search technique
\cite{Helst68,OwenS99}, particularly suitable to search for signals that
are characterized by a large number of wave cycles within the
interferometer observational window ($\simeq 40\,{\rm Hz} - 2$\,kHz), and
whose waveforms can be accurately modeled from a theoretical point of
view.  A well-known example of such a signal is the binary inspiral of
compact objects, either neutron stars (NS) or stellar-mass black holes
(BH).  Such inspiral events are the most promising candidates for
ground-based interferometers, and, for the instruments of first
generation, especially those with high total mass (BH/NS and/or BH/BH
binary systems), see \cite{CutleT02} and references therein.

A crucial component in the successful implementation of matched-filtering
is the construction of a reliable family of gravitational waveforms to be
used as ``templates''. As one might expect intuitively, the closer the
templates are to the ``true'' signal, the higher the signal-to-noise ratio
(SNR) at which a detection can be achieved. For this reason in recent
years a tremendous effort has been devoted to the computation of the
waveforms that characterize the inspiral phase, using different
approaches, such as post-Newtonian expansions (see \cite{Blanchet02} and
references therein for an extensive review) and Pad\'e approximants
\cite{DamouIS98,DamouIS00}.

The simplest family of templates corresponds to GW signals from two
point-masses orbiting each other.  However, it is possible that the true
inspiral is characterized by other more complicated effects, such as
relativistic spin-orbit and spin-spin couplings due to the presence of
high spins misaligned one with respect to the other and to the orbital
angular momentum. One of the most dramatic effects induced by such
coupling is the precession of the binary orbital plane around a fixed
direction, with a significant number of precession cycles within the GW
frequency band of interest. As a result of this precession the
polarization of the waves impinging on the detector changes during the
observation, which in turn produces an amplitude and phase modulation of
the signal at the detector output \cite{ApostCST94}. It is well known that
if such modulation is not accounted for in the search templates the
detection efficiency could decrease significantly, especially for binaries
containing BHs and/or characterized by large mass ratios
\cite{Apost95,Apost96}. This would lead to a reduction of the volume of
the Universe accessible by any given search and hence of the rate of
detection. Current predictions of the detection rates for binary inspiral
with the first-generation of laser interferometers are relatively low, in
the most optimistic cases up to just a few events per year even for
massive binaries with two black holes \cite{KalogB01,BelczKB02}. Such
estimates are obtained by assuming perfect match between the signal and,
at least, one of the templates of the filter bank. If modulations induced
by spin-orbit precession are not included into the family of templates,
the actual signal-to-noise ratio that can be achieved in any given search
is lower. If such a loss of SNR occurs for a significant portion of the
parameter space that characterize binary systems, the chance of detections
by LIGO and other ground-based interferometers would be severely
compromised. This would in turn require to construct a new family of
template that takes precession effects into account. In the case of
binaries with two neutron stars, both observations of pulsar spin periods
\cite{TayloW82,Wolsz91} and our current theoretical understanding of
neutron star tidal evolution \cite{BildsC92,Kocha92} suggest that the
spins are essentially negligible in neutron star binary systems.
Therefore, we concentrate on binaries with either a black hole and a
neutron star or two low-mass black holes.

The PN equations of motion, up to second post-Newtonian order, including
spin-orbit and spin-spin corrections, have been derived in \cite{Kidde95}.
However, in this paper, instead of integrating the complete set of
equations, we use an analytical approximation of the 1.5PN order equations
derived by Apostolatos {\em et al.} \cite{ApostCST94,Apost95}. This
approximation is valid in the regime of {\em simple precession}, which is
relevant in the following two cases~:  (i) two spinning objects of equal
mass $\l(m_1=m_2\r)$ or (ii) unequal-mass objects but with only the most
massive one spinning $\l(S_2=0\r)$. Here we restrict our computations to
systems for which only the most massive object is spinning. In the case of
binaries with a black hole and a neutron star, such an assumption is
physically well justified based on the same arguments relevant for binary
neutron stars. In the case of binary black holes, the most massive of the
two is also expected to be spinning more rapidly, since it is the one that
in most cases is formed first, and therefore may have been spun up through
accretion from its non-degenerate companion, the progenitor of the second
black hole \cite{BelczKB02}. We restrict ourselves to systems with
relatively moderate total mass (i.e.  $m_1+m_2 \lesssim 10 M_\odot$) so
that one expects the PN-expansions to be fairly accurate \cite{BradyCT98}.
The detectability of two, high mass, non-spinning black holes, for which
the PN-expansions break down, has been recently assessed in
\cite{BuonaCV02}. Then, we consider whether the inspiral detection 
efficiency could be
improved by using a new family of ``mimic'' templates, initially suggested by
Apostolatos in \cite{Apost96} as a possible solution for the poor fitting
factor achieved by non precessing templates. This new class of templates
depends only on a small number of additional parameters -- three in this
case -- but attempt to ``mimic'' the effect of precession. Apostolatos
presented heuristic arguments supporting the choice of waveform: here we
quantitatively explore for the first time whether they indeed recover most
of the signal to noise ratio which is lost due the modulation effects. The
result of this investigation is actually that this new family of
templates is still inadequate to search for precessing binaries over
essentially the entire parameter space. 

This paper is organized as follows: Sec. \ref{s:signal} presents briefly
the simple precession regime and reviews the various types of templates
used throughout this paper, (post)-Newtonian and the ``mimic'' templates
which include a
correction term to the phase. The concept of the fitting factor is also
introduced. Results are shown in Sec. \ref{s:results}. We draw conclusions
and discuss lines for future work in Sec.  \ref{s:conclu}.

\section{Adopted signals and templates}\label{s:signal}

 \subsection{Simple precession formalism}

Apostolatos et al.~\cite{ApostCST94} have obtained a semi-analytical
approximation of the waveforms to 1.5PN order, called {\em simple
precession}. This approximation applies to a wide range of astrophysically
relevant scenarios and is adopted here.  The key effect of precession is to
change the orientation of the orbital plane during the observation, and
therefore introduce an amplitude and phase modulation in the signal
recorded at the detector output due to the time-varying response of the
detector to the two different polarizations of the GW. The effects of
precession give rise to the following waveform~:
 \be
 \label{e:Lambda}
 \tilde{h}_{\rm prec}\l(f\r) = \Lambda\l(f\r) \tilde{h}\l(f\r) 
= {\rm AM} \l(f\r) {\rm PM}\l(f\r)\tilde{h}\l(f\r).
 \ee
 $\tilde{h}\l(f\r)$ is the waveform without precession, $\Lambda\l(f\r)$
represents the modulation factor of the waveform due to the change of
relative orientation of the detector with respect to the binary orbital
plane. $\Lambda$ can be decomposed as an {\em amplitude modulation} ${\rm
AM}$ and a {\em phase modulation} ${\rm PM}$. The simplicity of this
regime is related to the fact that there is an analytical solution for
both ${\rm AM}$ (Eq. (11) of \cite{Apost95}) and ${\rm PM}$ (Eq. (14) 
of \cite{Apost95}). More details can be found in \cite{ApostCST94,Apost95}, but 
here we only outline the key features of the simple precession regime. We focus on
binaries where only the most massive object ($m_1$) carries a spin
($S=S_1$). As explained in the introduction this case is physically the
most relevant based on our astrophysical understanding of the formation of
binary compact objects \cite{BelczKB02}.

The key feature of simple precession is that the direction of the total
angular momentum ${\bf J}$ remains essentially constant throughout the
whole inspiral; the orbital momentum ${\bf L}$ and the spin ${\bf S}$ are
locked together and precess at the same rate around ${\bf J}$ with
constant tilt angle.  We also note that throughout our calculations, as a
first step, we do not include the {\it Thomas precession phase} (Eq.\ (14) 
in \cite{Apost95}).

Even though the expressions of the modulations ${\rm AM}$ and ${\rm PM}$
are not given in this paper, it is necessary to review the parameters on
which they depend. The position and orientation of the binary with respect
to the detector depends on four angles. The position of ${\bf L}$ on its
precession cone involves another angle
(see Fig. (4) of \cite{ApostCST94} for example). Notice that those five angles are
randomly distributed. One also needs to give the constant magnitude of the
spin $S$ and the cosine of the misalignment angle $\kappa$. The last two
parameters on which the modulation $\Lambda$ depends are the two masses
$m_1$ and $m_2$. To summarize, the modulation function depends on nine
parameters, five of them being random angles. This large number of
parameters needed to model the precession modulation makes it very
unlikely, in the foreseeable future, to use the full precessing waveforms
as templates for matched filtering, due to the enormous computational
burden involved in such a search.

\subsection{Non-precessing waveforms}

 The simplest way of describing the binary inspiral without precession, is
to treat the objects as point masses in Newtonian gravity, and to
determine the radiation by the quadrupole formula. By doing so, and
applying the stationary phase approximation to transform the signal from
the time to the frequency domain, one obtains the well known {\em chirp
signal} \cite{SathyD91,DhuraS94,BalasD94}~:
 \be
 \label{e:chirp}
 \tilde{h}_0\l(f\r) = {\mathcal A} f^{-7/6} \exp \l(i \phi_0\l(f\r)\r)
\quad ; \quad
 \phi_0\l(f\r) = 2\pi t_c f - \phi_c -\frac{\pi}{4} +
 \frac{3}{128}\l(\pi{\mathcal M}\r)^{-5/3} f^{-5/3}.
 \ee
 In Eq. (\ref{e:chirp}), ${\mathcal M} \equiv
\l(m_1^3m_2^3/\l(m_1+m_2\r)\r)^{1/5}$ is the chirp mass, $t_c$ is the
coalescence time, $\phi_c$ the phase at coalescence and ${\mathcal A}$ a
constant amplitude, which depends on the source distance, mass, location
in the sky and orientation of the orbital plane. The Newtonian
non-precessing waveforms constitute a 3-parameter family~:~$t_c$, $\phi_c$
and ${\mathcal M}$. For the searches, however, it has be shown that the
maximization over the parameter $\phi_c$ is analytical
\cite{Apost95,Apost96} and a maximization over only two parameters is
needed.

Post-Newtonian corrections to the phase of (\ref{e:chirp}) can also be
included. The exact expressions are given by Eqs. (3) of \cite{Apost96}.
The PN corrections remove the degeneracy of the two masses into the single
chirp mass parameter. The total number of parameters in this case is four:
$t_c$, $\phi_c$ and the two masses $m_1$ and $m_2$.

The waveform (\ref{e:chirp}) can either be used as non-precessing part of
the signals ($\tilde{h}$ in Eq. (\ref{e:Lambda})) or as templates. In the
latter case, when PN corrections are included, as the templates used for
the real searches do not include spin effects, we set both the spin-orbit
and spin-spin coupling terms to zero ($\beta=0$ and $\sigma=0$ in Eqs. (3)
of \cite{Apost96}).

Let us mention that it is crucial to maintain the same PN-order for the
non-precessing part of both the signal and the templates. Indeed, we want
to ensure that whatever decrease in detection rate we calculate is related
solely to precession effects and not from mismatch in the PN-order of the
non-precessing parts (this consistency requirement was not satisfied in
\cite{Apost96}). In the course of our calculations we check that the
inclusion of higher order corrections does not significantly modify the
results obtained when considering non-precessing parts of Newtonian order
only.

\subsection{``Mimic'' templates}\label{ss:mimic}
To alleviate the problem of the large number of template parameters,
Apostolatos \cite{Apost96} suggested the use of a new family of templates
that still depends 3 extra parameters (beyond those associated with the
non-precessing parts) and might be able to capture the main features of
the precession modulation. He provided heuristic arguments that tends to
suggest that this new class of templates could be effective in searching
for precessing binaries, but never carried out a detailed {\em
quantitative} analysis. The main effect of precession is to introduce
an oscillatory modulation, for both the phase and the amplitude of the
waveforms, which is completely different from the secular evolution of the
GW phase. As the technique of matched-filtering rely on the template to
stay in phase with signal over the frequency region where most of the SNR
is accumulated, phase corrections are more crucial than amplitude
corrections. As a first step in improving the filters, Apostolatos
\cite{Apost96} suggested the inclusion of an oscillatory term to the phase
$\phi$ of the templates:
 \be
 \label{e:correction_23}
 \tilde{h}_{\rm cor}\l(f\r) = {\mathcal A} f^{-7/6} \exp \l(i \phi\l(f\r)  
+ i\phi^{\rm cor}\l(f\r)\r) \quad ; \quad \phi^{\rm cor\l(-2/3\r)} =
{\mathcal C} \cos\l({\mathcal B} f^{-2/3} + \delta\r).
 \ee

$\phi$ denotes the phase of the non-precessing waveform. If we consider only
Newtonian order then $\phi$ is only $\phi_0$ of Eq. (\ref{e:chirp}).
We call the templates with such an additional term the {\em mimic
templates}, because they represent an attempt to reproduce the main
behavior of precession, and hopefully increase the detection rate.  The
mimic templates have three additional parameters ${\mathcal C}$,
${\mathcal B}$ and $\delta$, with the exponent of the dependence on the
frequency being fixed. The choice of the exponent $-2/3$ comes from
approximate formulae for the precession angle (cf. Eqs(29) of
\cite{Apost95}). Indeed the value of the 
pulsation caused by precession relates closely to
the number of precession cycles. Originally the choice of the exponent
$-2/3$ was made by assuming that precession was only important for high
values of $S$. However,we will see that, as stated in \cite{Apost96}, the
choice of the exponent is not crucial and one can, instead of $-2/3$, use
the other approximate value $-1$ (cf. Eqs(29) of \cite{Apost95}): $
\phi^{\rm cor\l(-1\r)} = {\mathcal C} \cos\l({\mathcal B} f^{-1} +
\delta\r)$.

Considering only non-precessing parts of Newtonian order, we are left with
five parameters : ${\mathcal M}$, $t_c$ and the three additional
parameters ${\mathcal C}$, ${\mathcal B}$ and $\delta$. Performing the
maximization in the full 5-parameter space is beyond what our
computational resources allow.  Fortunately, a simplification stems from
the fact that the two chirp parameters ${\mathcal M}$ and $t_c$ are
essentially uncorrelated with those of the mimic correction
\cite{Apost96}.  This implies that, instead of doing a maximization over 5
parameters, one can first maximize over the two parameters ${\mathcal M}$
and $t_c$ and {\em then} over the last three. This approach greatly
reduces the computational time.

\subsection{The fitting factor}

Matched-filtering techniques are based on correlating the
output of the detector with a family of pre-calculated templates. The
highest possible SNR is then achieved when the signal $W$ coincides with
one of the templates. In practice, the signal is unknown and one has to
perform a search using a family of templates $T$ that depend on a given
set of unknown parameters $\lambda_1, \lambda_2,\ldots$. Let us denote
such a family with $T_{\vec{\lambda}}$. The fact that the exact signal is 
not included in the template family implies a loss in SNR
 \be
 \label{e:decrease_FF}
 \l(\frac{S}{N}\r) = {\rm max}_{\vec{\lambda}}\l[
\frac{\l(W|T_{\vec{\lambda}}\r)}
{\sqrt{\l(T_{\vec{\lambda}}|T_{\vec{\lambda}}\r)}}\r] = {\rm FF} \times
\l(\frac{S}{N}\r)_{\rm max} ={\rm FF} \times \l(W|W\r)^{1/2},
 \ee
 where $\l(\cdot|\cdot\r)$ is the scalar product of two waveforms,
weighted by the noise of the detector (Eq. (2.3) of \cite{CutleF94} ; see also
\cite{Apost96,Helst68,Apost95,Finn99} and references therein for more
details). The maximum SNR is given by $\l(W|W\r)^{1/2}$. The {\it fitting
factor} ${\rm FF}$ represents the loss of signal-to-noise ratio due to the
fact that the signal and the templates do not exactly overlap (note that
$0 \leq {\rm FF} \leq 1$). By definition, the fitting factor is expressed
by~:
 \be
 \label{e:FF}
 {\rm FF} = {\rm max}_{\vec{\lambda}}\l[
\frac{\l(T_{\vec{\lambda}}|W\r)}{\sqrt{
\l(T_{\vec{\lambda}}|T_{\vec{\lambda}}\r) \l(W|W\r)}}\r].
 \ee

In all our work, contrary to \cite{Apost96,Apost95}, we use to noise curve
of the initial configuration of LIGO. The low-frequency cut-off (40 Hz) is
at higher frequency that for advanced LIGO (10 Hz), and therefore the
effects of precession may be reduced (fewer precession cycles in the band
of the detector).

An important feature of the fitting factor is that it is independent of
the absolute normalization of the signal, the templates and the noise.
Should the templates, the signal or the noise be multiplied by any
constant, the value of ${\rm FF}$ would not change. However such scalings
change the value of $\l(S/N\r)_{\rm max}$ and therefore of $S/N$. It
implies, for example, that the ${\rm FF}$ does not depend on the constant
${\mathcal A}$ of Eqs.\ (\ref{e:chirp}).

Since we are interested in results that are independent of the specific
values of the random angles, we decide to show the ${\rm FF}$ averaged
over a given number of configurations of the random angles (chosen from a
uniform distribution in solid angles). We note $S/N$ implies a reduction,
by the same amount, of the maximum distance at which a source can be
detected with a given signal-to-noise ratio. The total accessible volume
therefore decreases by a factor of $<{\rm FF}>^3$. Assuming that sources
are homogeneously distributed in space, then the detection rate is
decreased by that same factor $<{\rm FF}>^3$.

\section{Results}\label{s:results}

\subsection{Computational parameters}

The determination of the ${\rm FF}$ requires a maximization over the
template parameters. Such maximization could be done by using
sophisticated algorithms but it turns out that the dependency of ${\rm
FF}$ on the parameters is rather complicated (see Fig. 2 of
\cite{Apost95}). The presence of several local maxima could lead to
convergence difficulties.  In this paper we adopt a more straightforward
way, which we have tested exhaustively. For each parameter of the
templates, we determine a range of values needed for the search and
populate such interval with discrete values of the parameters. Then, the
${\rm FF}$ is computed for all sets of parameters, allowing us to
determine the maximum throughout the explored space. The smoothness of our
curves is evidence in support of our assessment that we do find the global
maxima.

For each parameter, the interval of search for the maximum is obtained by
a trial search, with a very large interval, for a case of strong
precession (cases of weaker precession require narrower ranges, but we
still use the limits determined by the worst cases of strong precession).
The number of templates used to populate this interval is calculated to
ensure that the accuracy of ${\rm FF}$ is always below $10^{-2}$
(typically $30-100$ per parameter), based on convergence studies.
Similarly we determine the appropriate number of collocation points
(1,000) for the computation of the cross-correlation integrals, and the
number of sets of random angles (2,000) used to perform the angle average.
We note that the grids of the chirp mass and the total mass are
logarithmic (because powers of the masses appear in the expressions for
the templates).

\subsection{Importance of precession for detection rates} 

We have calculated the angle-averaged fitting factor $<{\rm FF}>$ for
various sets of masses ($\l(3,3\r)$, $\l(6,5\r)$, $\l(7,3\r)$ and
$\l(10,1.4\r)$ solar masses) and for the full range of spin magnitude
(from 0.1 to 1 of the maximum value with steps of 0.1) and misalignment
angle (from -1 to 1 for the cosine of the angle $\kappa$ with steps of
0.2).  In favor of briefness, we present here the results for the cases of
strongest precession : $m_1 = 10M_\odot$, $m_2 = 1.4 M_\odot$ and for a
subset of spin parameters. Indeed, as expected, the effects of precession
are stronger for higher mass ratios \cite{Apost96}.

\begin{figure}
 \includegraphics[height=6.5cm]{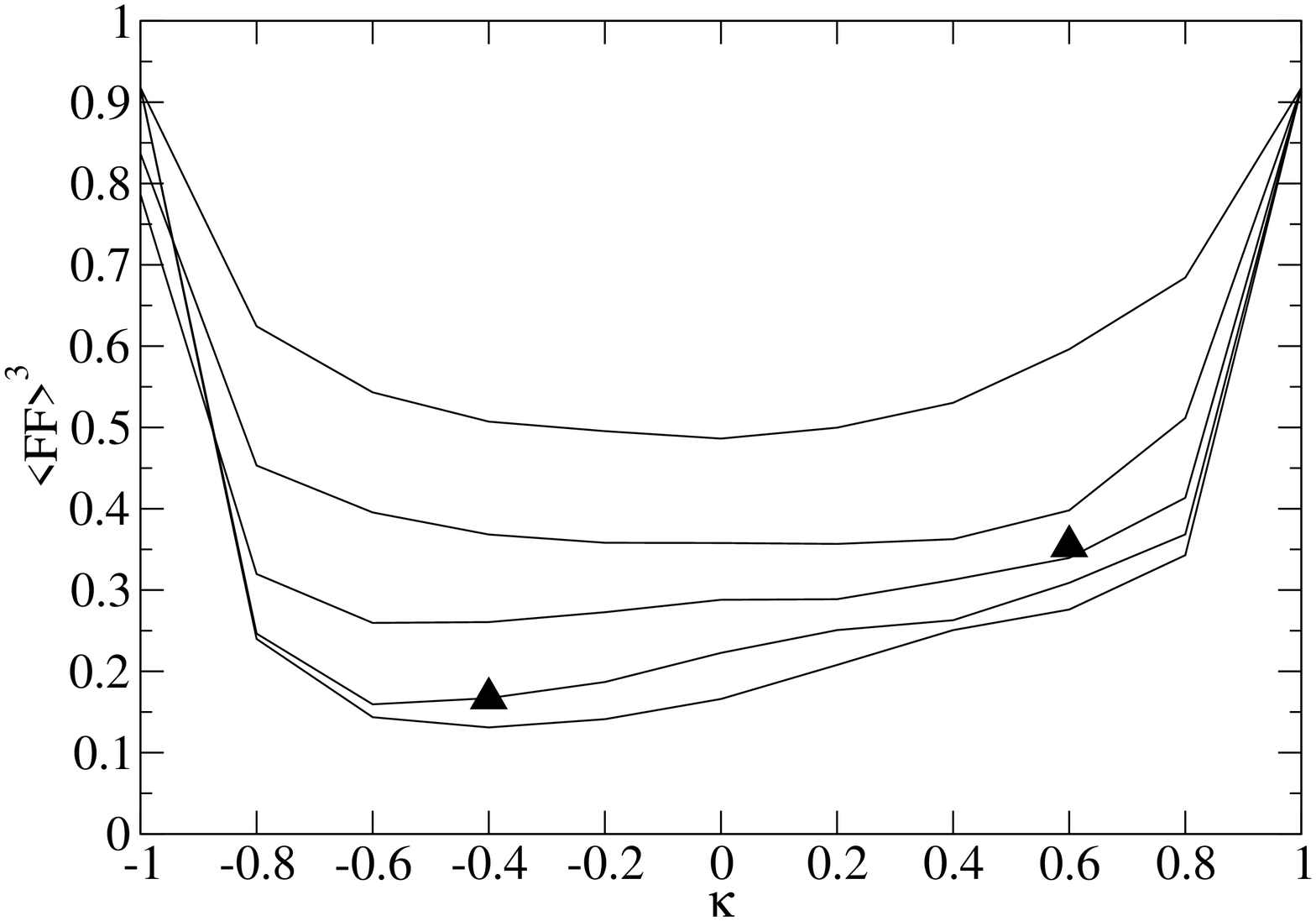}
 \includegraphics[height=6.5cm]{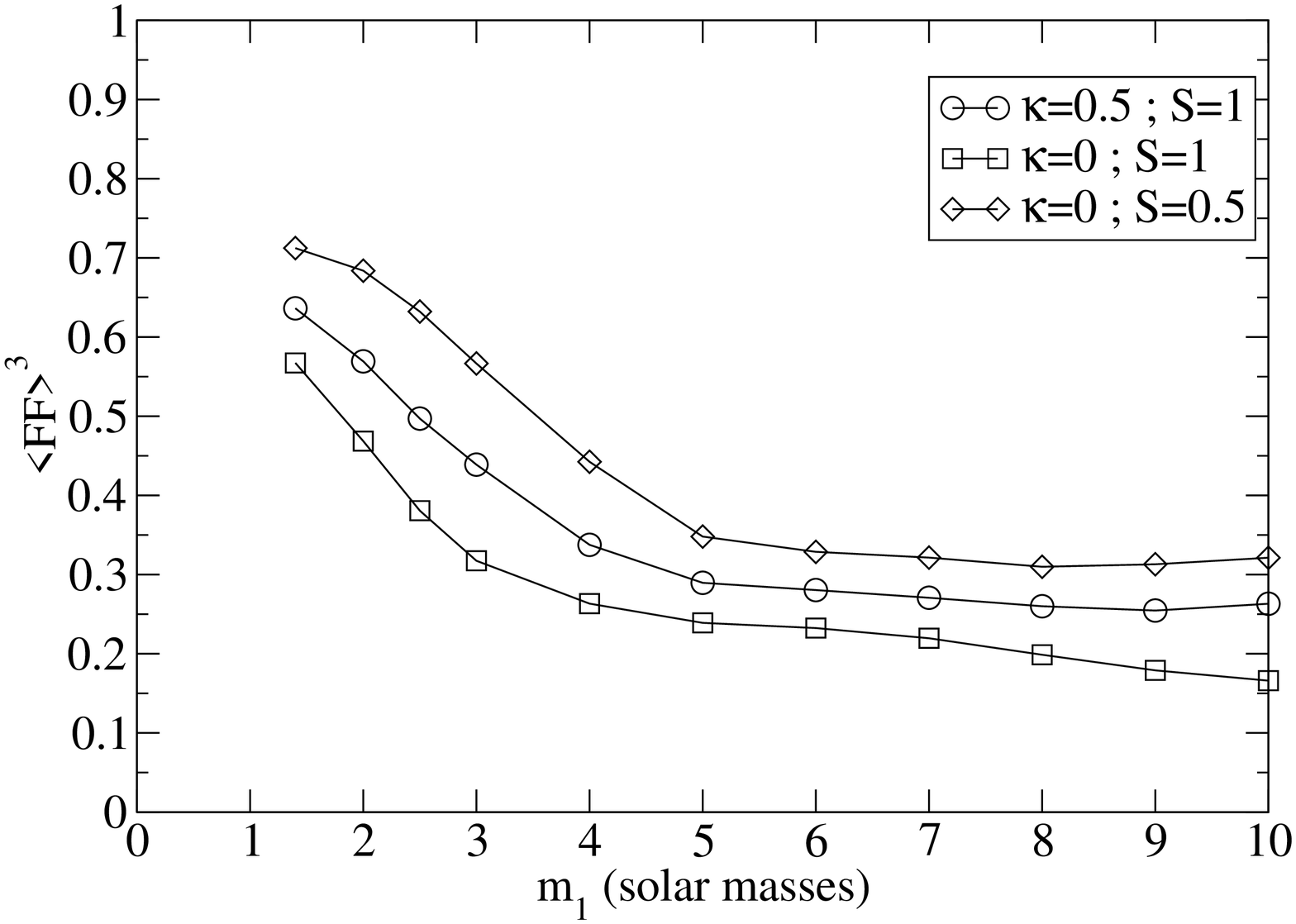}
 \caption{\label{f:newt} The left panel shows the value of $<{\rm FF}>^3$
as a function of $\kappa$, the cosine of the spin tilt angle. The masses
are $m_1 = 10 M_\odot$ and $m_2 = 1.4 M_\odot$. $S$ varies from $1$ (lower
curve) to $0.2$ (higher curve), with a step of $0.2$. The solid triangles
denote the results including 2PN corrections to the non-precessing parts,
for $S=1$. The right panel shows the dependence of the reduced detection
rate on the mass of the spinning object $m_1$. $m_2$ is fixed to
1.4\,M$_\odot$. Results are shown for 3 different sets of spin properties.}
 \end{figure}

The left panel of Figure \ref{f:newt} shows the reduction in detection
rate (i.e. $<{\rm FF}>^3$), as a function of $\kappa$, for various values
of the spin magnitude of the most massive object in the binary.  We have
also chosen a couple of specific cases to examine our initial conjecture
that PN effects would not change significantly the values of the $<{\rm
FF}>$, as long as they are added consistently in the signal and the
templates. The solid triangles show the rate reduction obtained when the
2PN correction to the phase of Eq. (\ref{e:chirp}) is included, for $S=1$.
In Figure \ref{f:newt} (left panel) we show the {\em largest} difference
we have found, while for all other mass pairs PN and Newtonian points
essentially overlap. Based on these small differences and our limited
computational resources, we have mostly used the Newtonian signals and
templates, for the non-precessing part of the waveforms. We note that for 
the other pairs of masses the maximum rate reduction lies in the range 
0.3--0.6. 

Figure \ref{f:newt} shows clearly that the detection rate could be very
significantly reduced, up to an order of magnitude. More precisely, the
detection rate drops by more than a factor of two for almost all values of
the spin magnitude, and virtually for all values of the misalignment angle
(i.e. all values of $\kappa$), even for angles as small as for $\kappa =
0.8$.  Given the low expected detection rates for BH-NS and BH-BH binaries
for the first-generation of ground-based detectors
\cite{KalogB01,BelczKB02}, if spin-induced modulations effects are not
included into the templates the chance of detection of BH-NS and BH-BH
binary systems is likely to be drastically reduced. This justifies the
quest for a new and more effective template family.

Our results can also be parameterized as a function of the mass of the
binary system. The right panel of Fig. \ref{f:newt} shows the variation of
$<{\rm FF}>^3$ with respect to $m_1$ (most massive, spinning object),
maintaining the other mass fixed at $m_2 = 1.4 M_\odot$. Three different
sets of spin properties are shown. Precession effects become more
important as $m_1$ increases, and therefore the number of precession
cycles within the relevant frequency band increases. As a result the
detection rate drops by factors of 5 to 10 for high values of $m_1$. One
can also see that the detection rate drops rather rapidly initially and
then saturates. Assuming that detection is possible as long as $<{\rm
FF}>^3$ is greater than a given threshold, say $0.5$, one can see that
binaries with $m_1 > 4M_\odot$ could remain undetected, showing once again
the need for templates that account for modulation effects.

\subsection{Efficiency of the ``mimic'' templates}

 \begin{figure}
 \includegraphics[height=6.5cm]{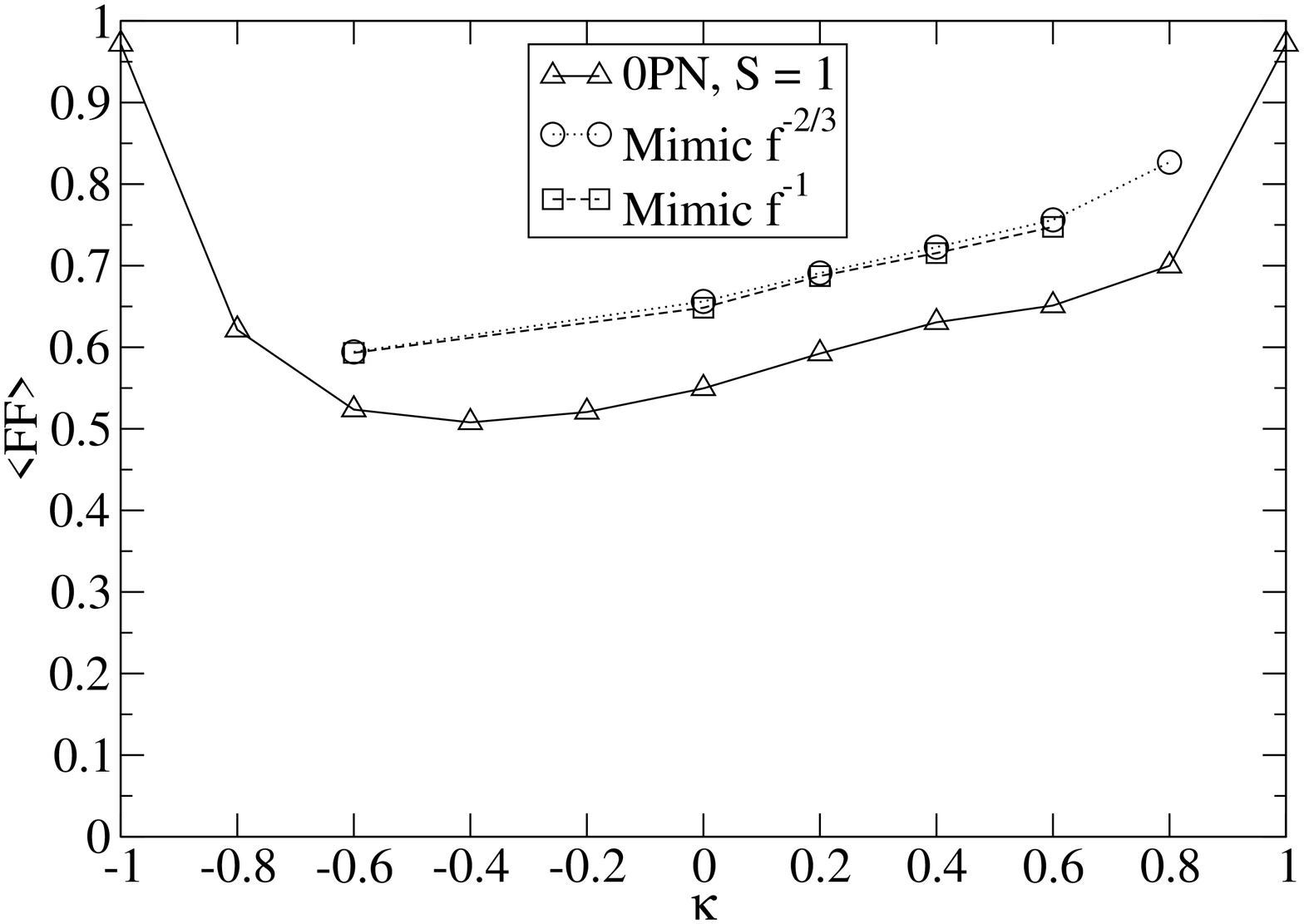}
 \includegraphics[height=6.5cm]{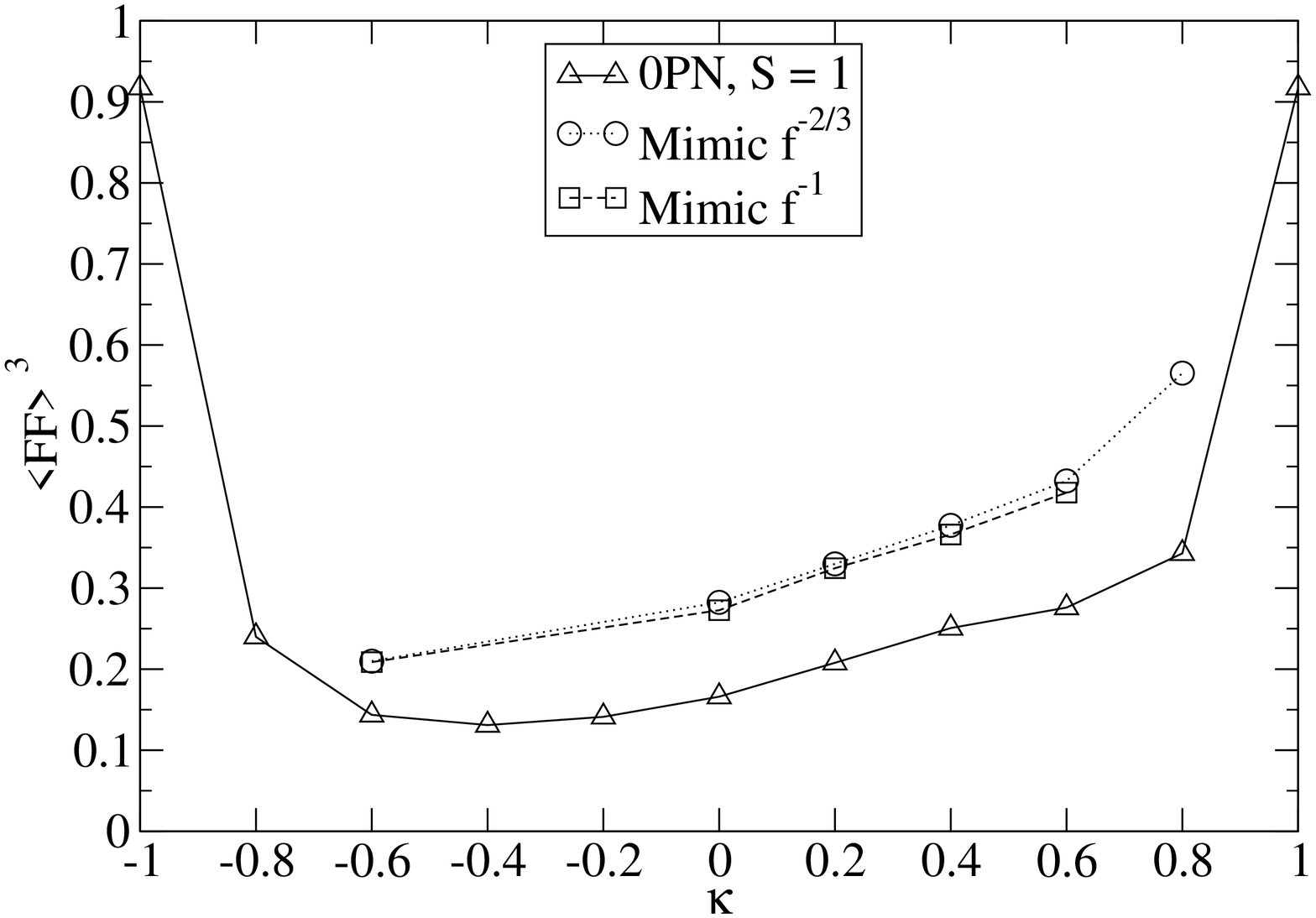}
 \caption{\label{f:mimic_10_14} Efficiency of the ``mimic'' templates
in recovering the signal-to-noise ratio and increasing the inspiral
detection rate. The triangles and solid lines correspond the results
obtained using the chirp templates without any precession modification
(\ref{e:chirp}).  The circles and dotted line denote results when the
phase correction is included with the $-2/3$ exponent, and
the squares and dashed line with the $-1$ exponent.  The left panel shows
the average $<{\rm FF}>$ whereas the right panel shows the reduction
factors of the detection rate, as a function of the cosine of the
spin-misalignment angle. The masses are $m_1 = 10M_\odot$ and $m_2 = 1.4
M_\odot$ and the spin magnitude is maximum ($S=1$).}
 \end{figure}

Figure \ref{f:mimic_10_14} shows the improvement in terms of $<{\rm FF}>$
(left panel) and $<{\rm FF}>^3$ (right panel) achieved by including
the ``mimic'' correction to the phase (Eq. (\ref{e:correction_23})). The
results are shown for two different choices for the exponent presented 
in Sec. \ref{ss:mimic}, and in comparison to results with
templates that do not include any precession-like modification.  As
mentioned above, the results do not vary significantly with the frequency
exponent~:~results obtained using the $-2/3$ and $-1$ are almost
identical. As expected there is an improvement of the fitting factor,
which is in the range 15-30\%.  Nevertheless, we are far from recovering a
high SNR. The detection rate can still reduced by up 80\%. 
For the other pairs of masses we have 
examined, fitting factors increase by about 10\%-20\% for the full range of 
spin properties. Based on these results we conclude that the ``mimic'' 
phase correction proposed by \cite{Apost96} 
alone is not enough to raise the detection efficiency to 
desirable levels.

\section{Conclusion}\label{s:conclu}
In this paper we present a systematic and quantitative study of the effect
of precession on the detection of binary inspiral signals.  Unlike earlier
work \cite{Apost96}, we focus on (i) implications for the initial LIGO
observations, (ii) the signal-to-noise ratio reduction due to the effects
of precession alone -- by analyzing signals and templates computed at the
same PN order for the non-precessing portion of the phase -- and (iii) the
quantitative study of the fitting factor of a family of mimic templates
that have been suggested for recovering high SNR.
We have examined results for 4 pairs of masses for the full range of spin 
properties and directions, and discussed in more detail the results of one 
of these pairs ($\l(10;1.4\r)$ solar masses) that exhibits the strongest 
precession effects and is appropriate for binaries with a black hole and a 
neutron star. 

We first addressed the question of how important is precession and for
what binary properties. We found that it can seriously affect detection,
if mass ratios are in excess of about 2, spin magnitudes in excess of
about 30\% of maximum, and spin tilts in excess of about 35 degrees. These
results help in the identification of the parameter space where ways of
improving the detection efficiency must be found. We found that the
detection rate can decrease by almost an order of magnitude if searches
are performed with templates that do not include precession effects. Such
a loss of events can be very concerning, given the current low-estimates
for the expected detection rates \cite{BelczKB02}.

Precession waveforms depend on a large number of parameters and their use
as a template family is not feasible. Therefore, it is important to
introduce a family of templates that can ``mimic'' precession effects well
enough, but they depend on a small number of parameters. Here we tested
one such family suggested in \cite{Apost96}, and found that, although they
do increase the detection rate, this increase is not significant enough to
raise the detection rate to desirable levels (leading to improvement of
more than a factor of a few for the case of strong precession). We 
conclude that other forms of ``mimic'' templates must be explored in the 
near future. Such an exploration is beyond the scope of the present paper,
but we are undertaking it as the next step in this project.

One can further restrict the physical parameter space over which
precession is crucial for detection, by convolving our results from the
first part of the paper with astrophysically relevant distributions of
binary and spin parameters of double compact objects, derived based on our
current understanding of the formation of compact binaries with black
holes. We are currently working on such a convolution with detailed
population calculations for many formation models \cite{BelczKB02} of BH
binaries, with the goal of obtaining an astrophysically-motivated picture
of how important it will be to include precession in the template families
for the search of GW inspiral signals in the next few years
\cite{IhmKGB02}.

\begin{acknowledgments}
 We would like to thank B.~Allen, P.~Brady, A.~Buonanno, and B.~Owen for
useful discussions. This work is supported by NSF Grant PHY--0121420. We
are also grateful to the High Energy Physics group at Northwestern
University for allowing us to access their computer cluster {\sc THEMIS}.
 \end{acknowledgments}

\end{document}